\documentclass[aps,prd,twocolumn,superscriptaddress,nofootinbib,showpacs,10pt]{revtex4-1}

\usepackage[left=2.0cm,right=2.0cm,top=2.0cm,bottom=2.0cm,marginparwidth=17mm]{geometry}
\usepackage[utf8]{inputenc}
\usepackage[english]{babel}
\usepackage{amsmath}
\usepackage{bbm}
\usepackage{amsfonts}
\usepackage{dsfont}
\usepackage{amssymb}
\usepackage{graphics}
\usepackage{graphicx}
\usepackage{subfigure}
\usepackage[usenames,dvipsnames,svgnames]{xcolor}  
\usepackage[linktocpage]{hyperref}   
\usepackage{todonotes}
\usepackage{ulem} 
\usepackage{braket}
\usepackage{comment}
\hypersetup{colorlinks=true,linkcolor=beamer@PRD, citecolor=beamer@PRD}

\renewcommand\eqref[1]{\textcolor{beamer@PRD}{(}\ref{#1}\textcolor{beamer@PRD}{)}}

\definecolor{beamer@PRD}{RGB}{46,48,146}
\begin{document}
\title{Enhanced quantum phase estimation with $q$-deformed nonideal nonclassical light}
\author{Duttatreya}
\affiliation{Department of Physics, Birla Institute of Technology and Science, Pilani, K K Birla Goa Campus, Zuarinagar, Sancoale, Goa 403726, India}
\author{Sanjib Dey}\email{sanjibd@goa.bits-pilani.ac.in}
\affiliation{Department of Physics, Birla Institute of Technology and Science, Pilani, K K Birla Goa Campus, Zuarinagar, Sancoale, Goa 403726, India}
\begin{abstract}
We investigate quantum phase estimation in a Mach–Zehnder interferometer using $q$-deformed photon states, including $q$-coherent and $q$-cat states, which model realistic deviations from ideal light sources. By deriving closed-form photon count likelihoods via the Jordan–Schwinger mapping, we compute the quantum and classical Fisher information and perform Bayesian inference on simulated detector data. Our results show that photon counting remains an optimal measurement strategy even for deformed states, with classical and quantum Fisher information in exact agreement. Furthermore, the phase sensitivity improves with increasing $q$-deformation, indicating enhanced metrological performance driven by nonclassical photon statistics. These findings highlight the utility of $q$-deformed states in quantum sensing.
\end{abstract}


\maketitle
\section{Introduction}
Breakthroughs in science and technology are frequently driven by advances in measurement precision, continually pushing the boundaries of what can be resolved at fundamental scales. Quantum metrology, which utilizes the principles of quantum mechanics to enhance measurement sensitivity, offers tools that can outperform classical approaches. This has enabled transformative applications across diverse fields, including optical atomic clocks \cite{Ludlow_etal_2015}, attometer-scale displacement detection in gravitational-wave observatories \cite{Abbott_etal_2016}, and nanoscale quantum imaging \cite{Degen_Reinhard_Cappellaro_2017}. At the heart of many of these advancements lies the task of phase estimation, a central technique that determines an unknown phase parameter encoded in a quantum probe from measurement outcomes \cite{Giovannetti_Lloyd_Maccone_2011}.

A classic setup for optical phase estimation is the Mach–Zehnder interferometer (MZI), a unitary $SU(2)$ device where an unknown phase-shift $\phi$ is imparted onto one optical path and inferred via interference at a final beam splitter~\cite{Caves_1981}. The MZI is a simple and loss-resilient setup that is widely used, ranging from large-scale systems like LIGO to compact integrated photonic circuits. The standard implementations using coherent laser input and intensity measurements achieve sensitivity at the standard quantum limit (SQL), where the uncertainty scales as $\Delta\phi\propto 1/\sqrt{N}$, with $N$ being the average number of photons. To surpass SQL, one typically resorts to either enhanced measurement schemes or nonclassical resources (probes) that embed greater information about the phase.

On the measurement side, photon-number-resolving detections have emerged as both practical and powerful. Once suitable detectors are available, such measurements are not only experimentally feasible but also informationally complete, enabling full reconstruction of the likelihood function $P(n_{1},n_{2}|\phi)$. Bayesian updating of this likelihood offers a natural and data-efficient pathway to optimal estimators and has been shown to saturate the quantum Cram\'er–Rao bound (QCRB) for coherent and squeezed light inputs both theoretically and experimentally~\cite{Pezze_etal_2007,Higgins_etal_2007,Ben-Aryeh_2012}. Accordingly, Bayesian inference turns out to be a natural choice for optimal and data-efficient estimation.

On the probe-state side, a range of nonclassical states has been explored for enhanced phase sensitivity surpassing the SQL. Among them, N00N states can, in principle, achieve the Heisenberg-limited (HL) sensitivity $\Delta\phi \propto 1/N$, but their extreme fragility to loss and low scalability (with demonstrations reaching only up to $N = 5$ photons) restrict their practical use~\cite{Nagata_etal_2007,Afek_Ambar_Silberberg_2010}. As an alternative, optical Schr\"odinger cat states, which are coherent superpositions of the form $\ket{\text{cat}_\pm}\propto\ket{\alpha} \pm \ket{-\alpha}$ offer improved scalability and enhanced robustness to loss, as their phase information is encoded within a single mode rather than through path entanglement~\cite{Munro_Spiller_2011,Ourjoumtsev_etal_2007,Deleglise_etal_2008}. Recent studies have shown that even in lossy environments, engineered high-photon-number HHG-cat states retain metrological utility~\cite{Stammer_etal_2024}.

However, real-world implementations rarely produce perfect coherent or cat states. Fluctuations, nonlinearities, and dissipative effects introduce deviations that cannot always be accurately captured by the standard bosonic algebra. This motivates the exploration of $q$-deformed oscillator algebras, where the canonical commutation relations are generalized to include a deformation parameter $q \in (0,1]$. $q$-deformed algebras that were first introduced in the context of quantum groups and mathematical physics~\cite{Arik_Coon_1976,Biedenharn_1989,Macfarlane_1989,Sun_Fu_1989,Kulish_Damaskinsky_1990}, have been shown to effectively model nonideal photon statistics, including sub- and super-Poissonian distributions encountered near lasing thresholds~\cite{Katriel_Solomon_1994,Dey_etal_2013, Dey_2015, Dey_Hussin_2016, Berrada_Eleuch_2019}.

The coherent states arising from these deformations denoted as $\ket{\alpha}_q$, exhibit tunable nonclassical features depending on $q$~\cite{Dey_2015, Dey_Hussin_2016, Dey_Hussin_2016}, and their superposition, $q$-deformed cat states have rich quantum properties including enhanced squeezing, sub-shot-noise characteristics, and mode entanglement~\cite{Dey_2015, Dey_Fring_Hussin_2017}. Despite their intriguing nonclassical nature, the metrological potential of $q$-deformed states has remained largely unexplored. In particular, questions about whether photon counting remains optimal and whether the phase sensitivity improves or degrades under deformation have not been systematically addressed.

In this work, we fill this gap by analyzing the performance of an MZI seeded with $q$-deformed coherent and cat states. Using the Jordan–Schwinger representation of the interferometric transformation, we derive analytical expressions for the photon count probabilities $P(n_{1},n_{2}|\phi,q)$, and compute both quantum and classical Fisher information. Bayesian inference is employed to simulate the estimation performance. Our key findings are at least threefold. First, the classical and quantum Fisher information are equal for all tested values of $q$, which means that the photon-number-resolving detection saturates the QCRB, confirming it as an optimal measurement for these deformed inputs. Second, the posterior variance obtained via Bayesian inference saturates the QCRB over a broad range of mean photon numbers. Third, for low photon numbers, phase sensitivity improves with increasing deformation, driven by the broadened photon-number distributions inherent in the $q$-deformed states. This improvement reflects the phase-number uncertainty relation $\Delta\phi\Delta n \geq 1$, where an increased number variance enhances phase precision. These results indicate that the $q$-deformed states serve as nonideal yet tunable quantum resources for quantum-enhanced metrology, which is capable of achieving quantum-limited precision, surpassing that of standard cat and coherent states, using only photon-number-resolving detectors and without the need for adaptive feedback. 

The remainder of this article is structured as follows. In Sec.\,\ref{sec2}, we introduce the $q$-oscillator algebra, define the probe states, outline the MZI model, and derive photon count probabilities. Sec.\,\ref{sec3} provides a detailed method of calculating the quantum Fisher information (QFI) for our states. Sec.\,\ref{sec4} describes the Bayesian inference procedure and simulation setup. In Sec.\,\ref{sec5}, we analyze the QCRB saturation and Fisher-information equivalence, followed by a detailed discussion of our key results. Finally, Sec.\,\ref{sec6} concludes our study and provides an outlook.
\section{Deformed nonlinear photon states} \label{sec2}
We commence with the construction of $q$-deformed nonlinear photon states from a one-dimensional $q$-deformed oscillator algebra~\cite{Arik_Coon_1976,Biedenharn_1989,Macfarlane_1989,Sun_Fu_1989,Kulish_Damaskinsky_1990,Dey_etal_2013,Dey_2015,Dey_Hussin_2016}
\begin{equation}
\hat{a}_q \hat{a}_q^\dagger-q \hat{a}_q^\dagger \hat{a}_q=\hat{\mathds{1}}, \quad |q|<1,
\end{equation}
which reduces to the Heisenberg-Weyl algebra  $[\hat{a},\hat{a}^\dagger]=\hat{\mathds{1}}$ in the limit $q\rightarrow 1$. The $q$-deformed ladder operators are defined by
\begin{alignat}{1}
& \hat{a}_q^\dagger \ket{n} = \sqrt{[n+1]_q}\ket{n+1} , \qquad \hat{a}_q\ket{0} : =0  \label{qLadder1}\\
& \hat{a}_q \ket{n} = \sqrt{[n]_q}\ket{n-1}, \qquad \langle 0 |0\rangle :=1, \label{qLadder2}
\end{alignat}
where $[n]_q=\frac{1-q^n}{1-q}$ denotes the $q$-deformed photon number. The so-called nonlinear coherent states~\cite{Filho_Vogel_1996, Manko_etal_1997, Sivakumar_2000, Roy_Roy_2000} are constructed as the eigenstates of the $f$-deformed annihilation operator, i.e., $\hat{a}_f\ket{\alpha,f} = \alpha \ket{\alpha,f}$, such that
\begin{equation}
\ket{\alpha , f}  = \frac{1}{\sqrt{\mathcal{N}_f}}\sum_{n=0}^{\infty}\frac{\alpha^n}{\sqrt{n!}g(n)}  \ket{n},
\end{equation}
where
\begin{alignat}{1}
& \hat{a}_f= \hat{a} f(\hat{n}), \quad \hat{a}_f\ket{n}= \sqrt{n} f(n)\ket{n-1} \label{fLadder1}\\
& \hat{a}_f^\dagger = f(\hat{n})\hat{a}^\dagger, \quad \hat{a}_f^\dagger \ket{n}= \sqrt{n+1} f(n+1) \ket{n+1}. \label{fLadder2}
\end{alignat}
Here, $\mathcal{N}_f$ is the normalization constant, $f(\hat{n})$ is an operator-valued function of the standard number operator $\hat{n}=\hat{a}^\dagger \hat{a}$, and 
\begin{eqnarray}
g(n) = \begin{cases}
1 & \text{if}~ n=0 \\
\prod_{m=0}^n f(m) & \text{if}~n>0.
\end{cases}   
\end{eqnarray}
The set of $q$-deformed ladder operators shown in \eqref{qLadder1}, \eqref{qLadder2} and $f$-deformed ladder operators in \eqref{fLadder1}, \eqref{fLadder2} become identical when
\begin{equation}
f(n)=\sqrt{\frac{[n]_q}{n}}.    
\end{equation}
For more information in this regard, see; for instance,~\cite{Dey_2015,Berrada_Eleuch_2019}. Thus, the $q$-deformed nonlinear coherent states~\cite{Katriel_Solomon_1994,Dey_2015,Dey_Hussin_2015, Berrada_Eleuch_2019} take the following form 
\begin{equation}\label{qNonlinearCoherent}
|\alpha\rangle_q=\frac{1}{\sqrt{\mathcal{N}_q}}\sum_{n=0}^\infty\frac{\alpha^n}{\sqrt{[n]_q!}}|n\rangle, ~~~ \alpha\in \mathbb{Z},
\end{equation}
where $\mathcal{N}_q=\sum_{n=0}^\infty\frac{|\alpha|^{2n}}{[n]_q!}$ represents the normalization constant. Unlike the Glauber coherent states, the nonlinear coherent states show nonclassical features~\cite{Roy_Roy_2000, Dey_Hussin_2015, Dey_Fring_Hussin_2017}. So, it is worth investigating the effect of the $q$-deformed nonlinear coherent states \eqref{qNonlinearCoherent} on phase estimation.

The $q$-deformed nonlinear cat states are built by superposing two coherent states \eqref{qNonlinearCoherent} of equal amplitude but different phases ~\cite{Mancini_1997, Dey_2015,Fakhri_Hashemi_2016}
\begin{equation}\label{qCat}
|cat\rangle_q^\pm=\frac{1}{\sqrt{\mathcal{M}_q}}\Big(|\alpha\rangle_q\pm|-\alpha\rangle_q\Big),
\end{equation}
where $ \mathcal{M}_q=2\pm\frac{2}{\mathcal{N}_q}\sum_{n=0}^\infty\frac{(-1)^n|\alpha|^{2n}}{[n]_q!}$. The states show exotic behaviors when the deformation parameter $q$ lies in the range $0<q<1$. In particular, the parameter $q$ can be tuned to vary the strength of the nonclassicality. For further details, see; for instance~\cite{Dey_2015}.

The real-life realization of nonlinear nonclassical states has been a central focus of quantum optics research, owing to their potential applications in quantum technologies \cite{Dodonov_2002_Review, Chang_Vuletic_Lukin_2014,Dey_Fring_Hussin_2018_Review, Dey_2021_Review}. These states, which arise from deformed algebras or effective nonlinear interactions, exhibit rich quantum features such as squeezing, antibunching, and quantum interference. Several theoretical works have proposed schemes for their generation using nonlinear Hamiltonians, atom-field interactions, or engineered deformation functions \cite{Wang_Goorskey_Xiao_2001, Naderi_Soltanolkotabi_Roknizadeh_2005, Lu_etal_2013, Yan_Zhu_Li_2016}. On the experimental front, significant progress has been made using platforms such as circuit QED, nonlinear optical media, and integrated photonics \cite{Gambetta_etal_2006, Solntsev_etal_2014, Latmiral_Mintert_2018, Kruk_etal_2019}. These advancements have enabled controlled generation and characterization of nonlinear nonclassical states, marking a promising step toward their integration in quantum-enhanced sensing, computation, and communication.
\begin{figure}
\includegraphics[width=.9\linewidth]{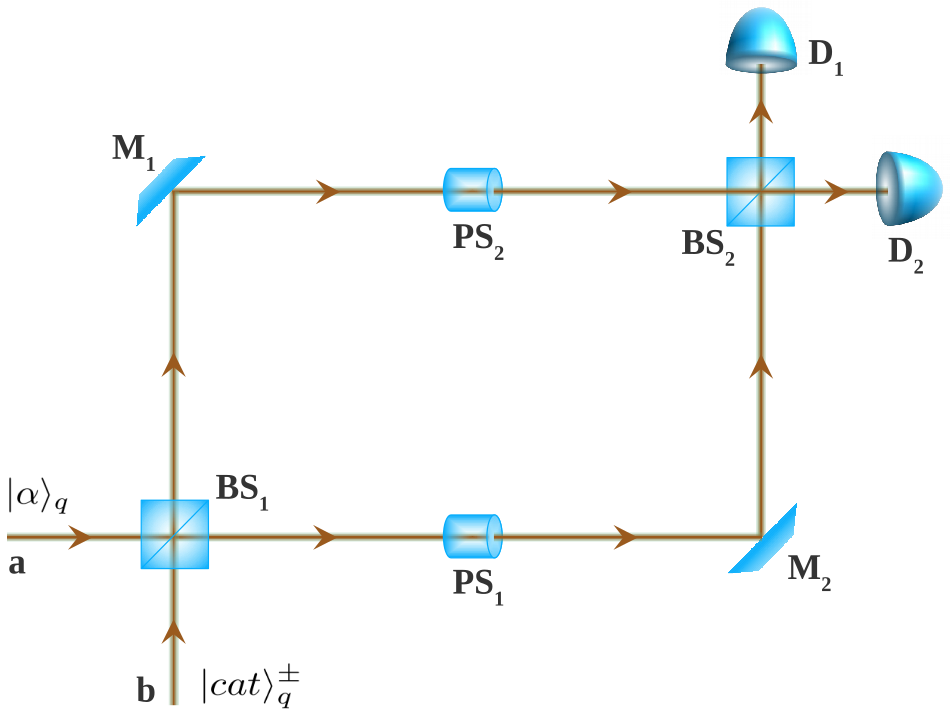}
\caption{(Color online). A schematic illustration of a Mach-Zehnder interferometric setup. Two pairs of input states $\ket{\text{cat}}_q^\pm$, $\ket{\alpha}_q$ are applied to the input modes of the first BS. A relative phase shift is applied by adjusting $\text{PS}_1$ and $\text{PS}_2$. Finally the output states of $\text{BS}_2$ are detected using the detectors $\text{D}_1$ and $\text{D}_2$.}
\label{FigMZI}
\end{figure}
\subsection{$|\alpha\rangle_q$ and $|\text{cat}\rangle_q$ through MZI}
To describe the action of the  MZI and its components, beam splitter (BS) and phase shifter (PS),  on the input states $|\alpha\rangle_q$ and $|\text{cat}\rangle_q$ we make use of the Jordan-Schwinger map that allows the mode operators to be written in terms of the angular momentum operators~\cite{Campos_Saleh_Teich_1989}
\begin{eqnarray}
&& \hat{J}_x=\frac{1}{2}(\hat{a}^\dagger \hat{b}+\hat{b}^\dagger a), \qquad \hat{J}_y=\frac{i}{2}(\hat{b}^\dagger \hat{a}-\hat{a}^\dagger \hat{b}), \nonumber\\
&& \hat{J}_z=\frac{1}{2}(\hat{a}^\dagger \hat{a}-\hat{b}^\dagger \hat{b}),     
\end{eqnarray}
which form a $SU(2)$ algebra $[\hat{J}_i,\hat{J}_j]=i\epsilon_{ijk}\hat{J}_k$ suitable for the mathematical description of a lossless MZI. Here, $(\hat{a},\hat{a}^\dagger )$ and $(\hat{b},\hat{b}^\dagger )$ represent the creation and annihilation operators of the first and second input modes, respectively, as shown in Fig.\,\ref{FigMZI}. In terms of the angular momentum operators $\hat{J}_x,\hat{J}_y,\hat{J}_z$, the action of a 50:50 BS is to rotate the input modes by an angle $\pi/2$ around the $x$-axis. In other words, the action of BS is equivalent to considering the BS operator $\hat{U}_{\textrm{BS}}=e^{-i\pi \hat{J}_x/2}$. PS rotates its input by an angle $\phi$ along the $z$-axis; thus, in a similar fashion, the PS is realized by the action of an equivalent PS operator $\hat{U}_{\textrm{PS}}=e^{-i\phi \hat{J}_z}$. As a whole, the MZI operation is equivalent to the MZI operator $\hat{U}_{\textrm{MZI}}=e^{-i\phi \hat{J}_x/2}e^{-i\phi \hat{J}_z}e^{i\phi \hat{J}_x/2}=e^{-i\phi \hat{J}_y}$. Therefore, the probability of getting clicks $n_1$ and $n_2$ at the detectors $D_1$ and $D_2$, respectively, conditioned on the induced  phase shift $\phi$ is given by
\begin{eqnarray}
p(n_1,n_2|\phi)&=&\Big|\langle n_1,n_2|e^{-i\phi \hat{J}_y}|\psi_{in}\rangle\Big|^2\nonumber\\
&=&\Bigg|\sum_{n=0}^N C_{N-n}K_n d^{N/2}_{\mu,\frac{N}{2}-n}(\phi)\Bigg|^2,
\end{eqnarray}
where $N=n_1+n_2$, $\mu=\frac{n_1-n_2}{2}$  and $d^j_{m^\prime,m}(\phi)$ are elements of the Wigner $d$-matrix~\cite{Wigner_1959_Book,Pezze_Smerzi_2008}. Here
\begin{equation}
C_n=\frac{\beta^n}{\sqrt{\mathcal{N}_q[n]_q!}}, \quad K_n=\frac{(e^{i\pi/2}\alpha)^n[1+(-1)^n]}{\sqrt{\mathcal{M}_q \mathcal{N}_q[n]_q!}},    
\end{equation}
where the phase $e^{i\pi/2}$ is introduced to fulfil the phase-matching condition~\cite{Liu_Jing_Wang_2013}.
\section{Quantum Fisher Information}\label{sec3}
The precision in estimating the phase for an unbiased estimator is bounded from below by the Cram\'er-Rao bound
\begin{equation}
\Delta\phi\geq\frac{1}{\sqrt{\nu F_Q}}, ~~~ 0\leq\phi\leq\pi,
\end{equation}
where $F_Q$ is the QFI, which can be calculated for the MZI setup using the expression
\begin{equation}\label{QFI}
F_Q=2\sum_{j,k}\frac{|\langle\lambda_j|\partial_\phi\rho_\phi|\lambda_k\rangle|^2}{\lambda_j+\lambda_k}.
\end{equation}
Here, $\rho_\phi$ is the density matrix of the output states of $\text{BS}_2$ and $\{|\lambda_j\rangle\},\lambda_j$ are the eigenstates and eigenvalues of $\rho_\phi$, repsectively. For pure states $\ket{\psi_\phi}$, the QFI \eqref{QFI} reduces to
\begin{equation}
F_Q=4\left(\langle\partial_\phi\psi_\phi|\partial_\phi\psi_\phi\rangle -|\langle\partial_\phi\psi_\phi|\psi_\phi\rangle|^2 \right),
\end{equation}
where $|\partial_\phi\psi_\phi\rangle=\frac{\partial}{\partial\phi}|\psi_\phi\rangle$. Since $|\psi_\phi\rangle=e^{-i\phi \hat{J}_y}|\psi_{in}\rangle$ is the output state of the MZI, the above equation modifies to
\begin{equation}
F_Q=4(\langle\psi_{in}| \hat{J}_y^2|\psi_{in}\rangle -\langle\psi_{in}| \hat{J}_y|\psi_{in}\rangle^2)\equiv4\triangle^2\hat{J}_y.
\label{eq:var}
\end{equation}
For the input states $|\alpha\rangle_q$ and $|\text{cat}\rangle_q$, we have $\langle\psi_{in}| \hat{J}_y|\psi_{in}\rangle=0$ and, thus, the exact expression of the QFI \eqref{eq:var} turns out to be 
\begin{alignat}{1}
F_Q &= \sum_{n,m} |C_n|^2 |K_m|^2 (n+ m+2mn) - (K_m K_{m+2}^* C_n^* C_{n+2} \nonumber\\
&+ K_m^* K_{m+2} C_n C_{n+2}^*) \sqrt{(n+1)(n+2)(m+1)(m+2)}. \label{QFIFInal}
\end{alignat}

\begin{figure*}[ht]
  \subfigure[]{
\includegraphics[width=.48\textwidth]{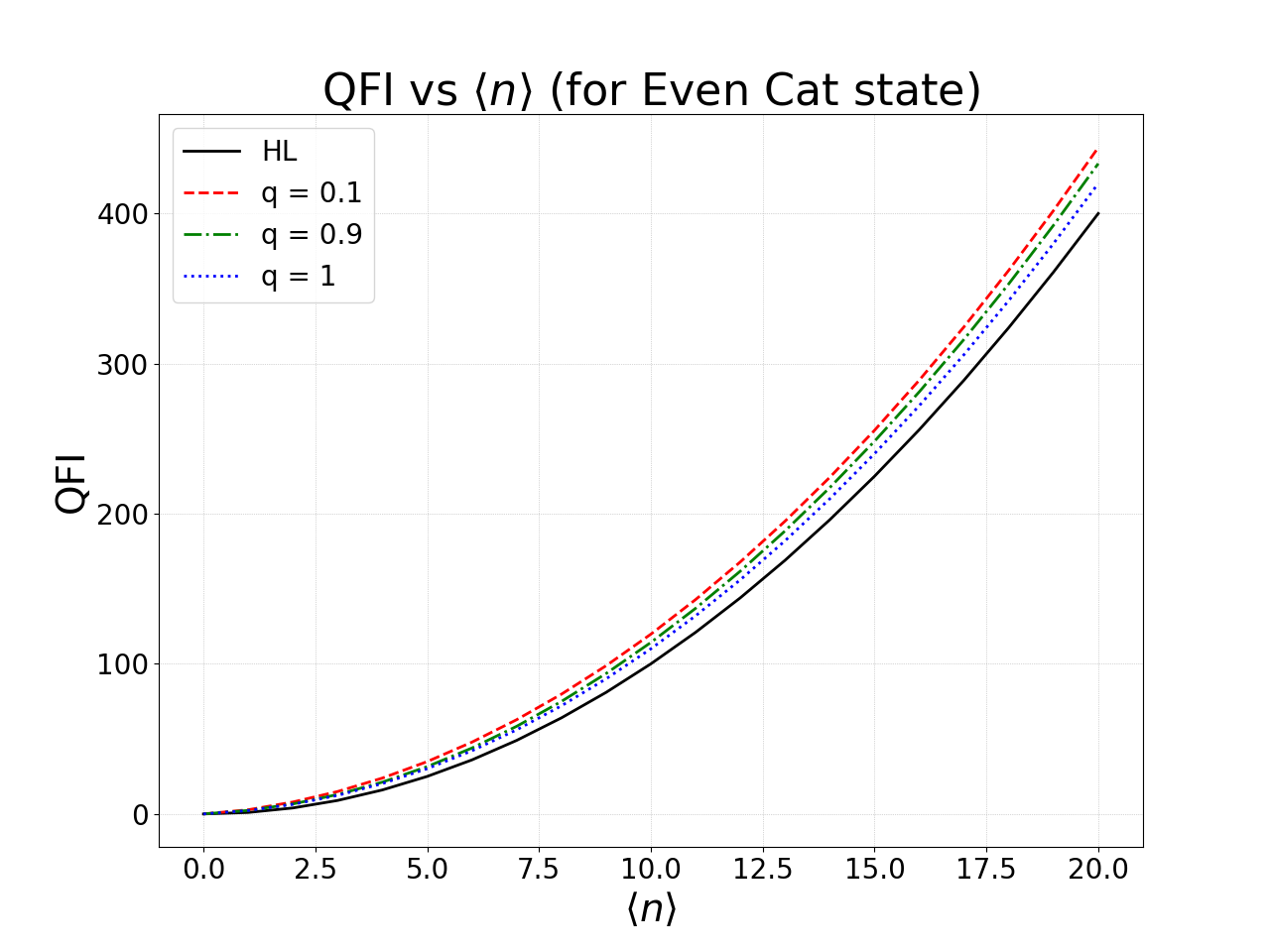}
\label{fig2a}}
\subfigure[]{
  \includegraphics[width=.48\textwidth]{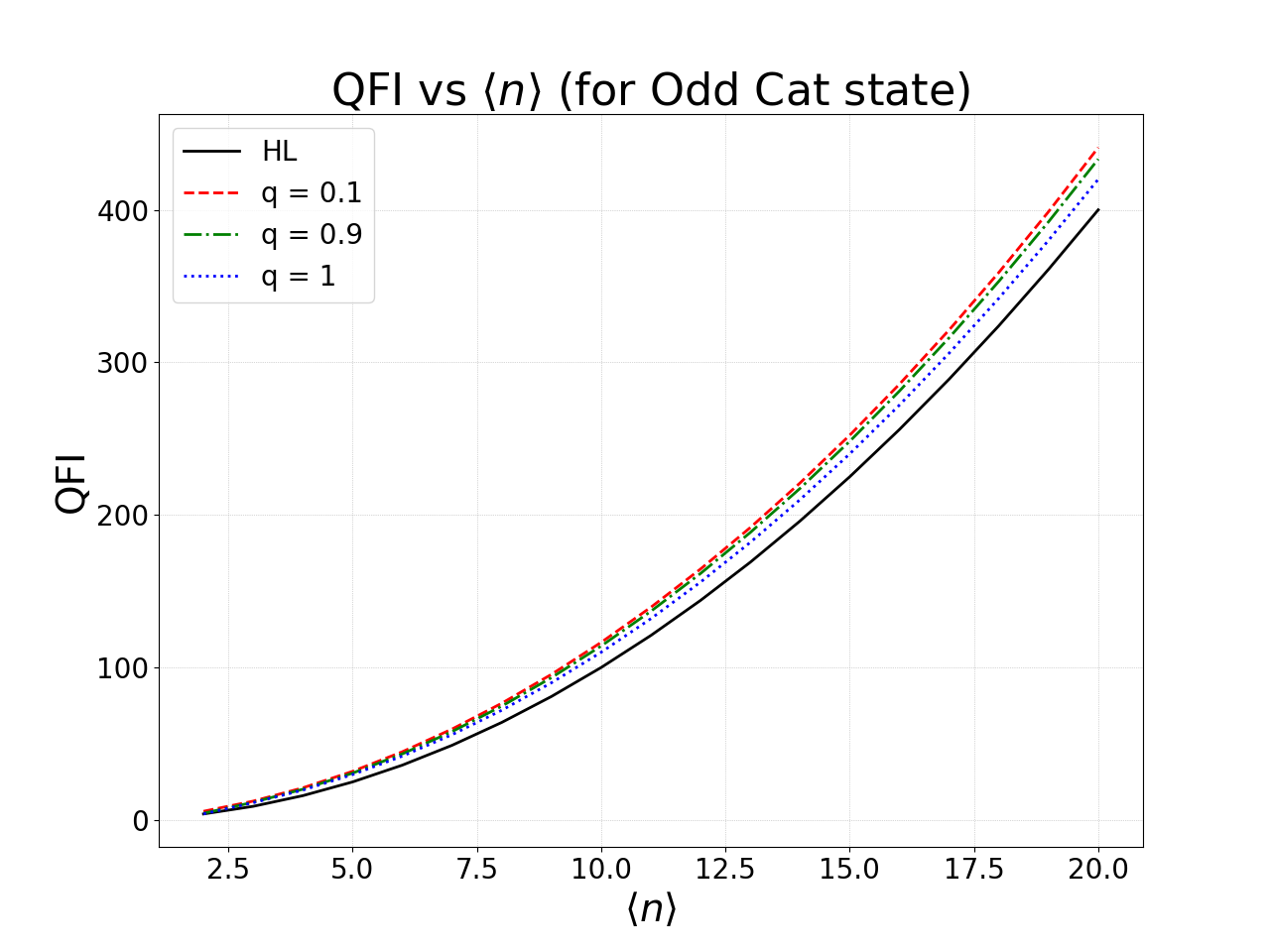}
  \label{fig2b}}
 \caption{(Color online). QFI ($F_Q$) versus the average input photon number $\langle \hat{n}\rangle$ for two different values of the deformation parameter $q$ represented by the red (dashed) and green (dashed-dot) lines. The blue (dotted) line corresponds to the undeformed (usual) cat and coherent state input, while the black (solid) line represents the Heisenberg-limited case.  \subref{fig2a} $q$-deformed even cat state \subref{fig2b} $q$-deformed odd cat state. $n_{max} = 30$ was chosen for all computations.}
\label{fig:QFIvsn}
\end{figure*}
\section{Bayesian inference} \label{sec4}
It is known that the parity measurements do not saturate the QFI using MZI for lossy inputs~\cite{Munro_Spiller_2011}. Usually, the parity measurement works well for the lossless ideal states like N00N, bat, pair coherent states, etc.~\cite{Gerry_Mimih_2010, Cooper_Hallwood_Dunningham_2010, Wang_etal_2019}, but may not extract full information from the states in general scenarios. On the other hand, a statistical inference method uses prior distributions and measurement outcomes to update the probability distribution over the parameter. This method can approach the QCRB asymptotically if the measurement is chosen wisely~\cite{Pezze_Smerzi_2008, Paesani_etal_2017}. Though the Bayesian inference method involves intense computational work, it performs more efficiently for general quantum states, particularly the nonideal states that we are using. This is why we resort to the Bayesian inference method and show that the photon counting saturates the QFI even in the $q$-deformed condition. For this, we consider the classical Fisher information (CFI) $F_C$, which is deduced by utilizing the probabilities of detector counts $p(\phi|n_1,n_2)$ for the MZI setup
\begin{equation}
F_C=\sum_{n_1,n_2}^\infty \frac{1}{p(n_1,n_2|\phi)}\left(\frac{\partial{p(n_1,n_2|\phi)}}{\partial{\phi}}\right)^2.
\label{eq:fsh}
\end{equation}
It is verified numerically that  $F_C$ provided by the detector counts is equivalent to the expression of QFI provided in (\ref{eq:var}). This led us to conclude that the photon counting measurement is optimal. Using the Bayes' theorem and the CFI given in (\ref{eq:fsh}), we infer the probability of $\phi$ conditioned on clicks  $n_1,~n_2$ as 
\begin{equation}
p(\phi|n_1,n_2)=\frac{p(n_1,n_2|\phi)p(\phi)}{p(n_1,n_2)}.
\end{equation}
Thereafter collecting the data of clicks $(n_1^{(\nu)},n_2^{(\nu)})$ from $\nu$ independent measurements, the probability distribution for $\phi$ is deduced as
\begin{equation}
p(\phi|((n_1^{(1)},n_2^{(1)}),\dots,(n_1^{(\nu)},n_2^{(\nu)}))=\prod_i^\nu p(\phi|n_1^{(i)},n_2^{(i)}).
\end{equation}
In the Bayesian phase estimation protocol, each experimental repetition enlarges the joint–count likelihood by one dimension, so the cost of updating and storing the posterior grows as $\mathcal{O}\!\bigl(S^{\,\nu}\bigr)$, where $S$ is the number of possible outcomes per shot.  
Although the statistical error decreases as $g(\nu)=1/(\nu F_{Q})$, the algorithmic overhead rises exponentially.  
A pragmatic optimum \(\nu_{\mathrm{opt}}\) can, therefore, be defined by minimising a composite figure of merit
\begin{equation}
C(\nu)=\operatorname{Var}(\hat{\phi})\;+\;\lambda\,S^{\,\nu},
\end{equation}
with $\lambda$ weighting the available computational resources. In practice, we choose the smallest \(\nu\) for which the posterior over $\phi$ first becomes \textit{unimodal}; beyond this point, additional shots compress the single peak only marginally, and the incremental precision gain is outweighed by the rapidly increasing runtime and memory requirements.  These unimodality and diminishing-returns criteria follow standard convergence diagnostics in Bayesian phase estimation~\cite{Neeve_etal_2025}.


\section{Analysis of QFI} \label{sec5}
\begin{figure}
    \centering
    \includegraphics[width=0.5\textwidth]{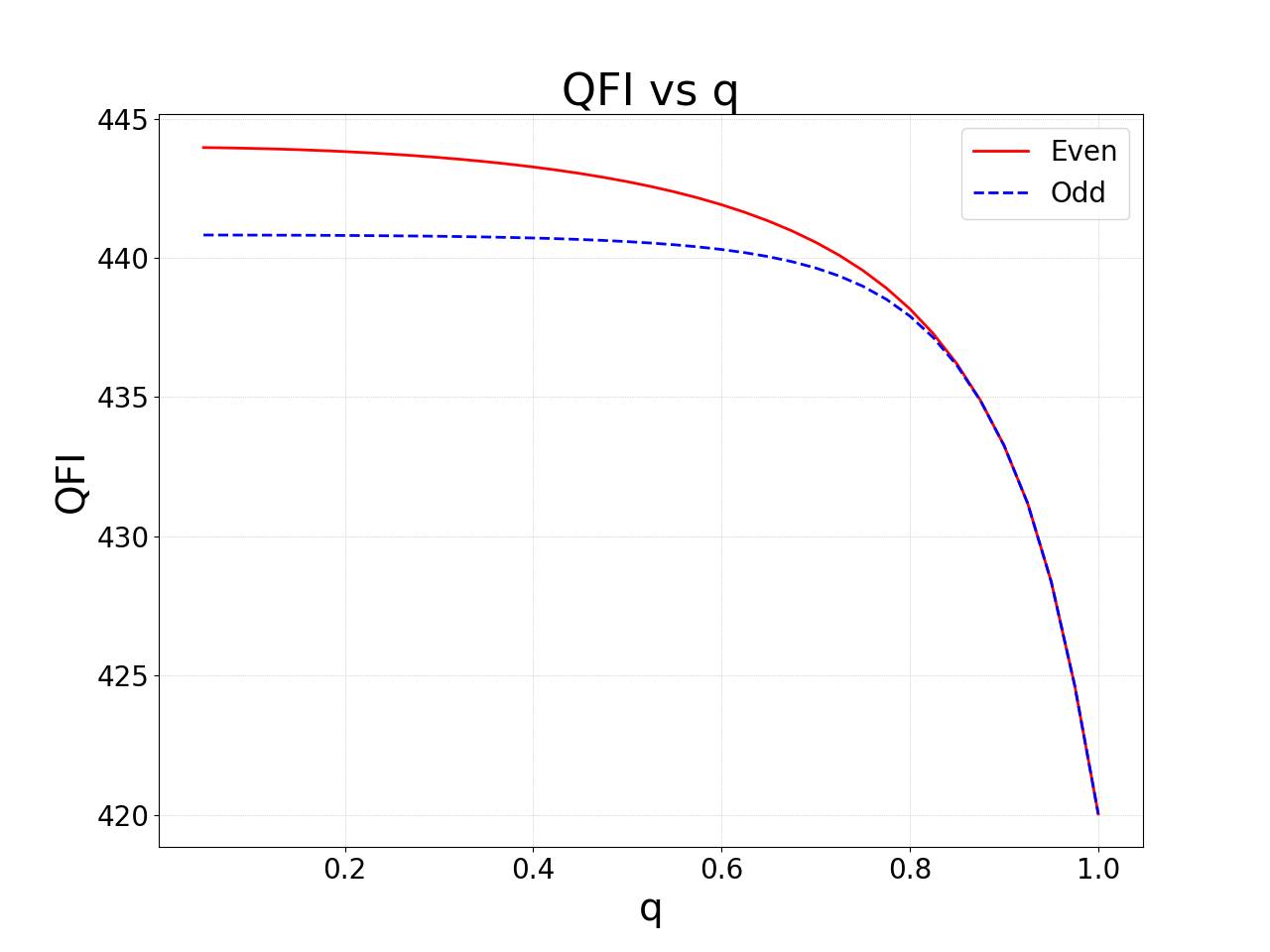} 
    \caption{(Color online). Variation of QFI with the deformation parameter $q$ for two different pairs of input states. The blue (dashed) line represents the behavior of the QFI for a $q$-coherent and an $q$-odd-cat state input, while the red (solid) line corresponds to a $q$-coherent and a $q$-even-cat input. In both cases, the expected photon number of the input is fixed at $\langle n \rangle = 20$. The QFI increases as the deformation parameter $q$ decreases, indicating that greater deformation leads to enhanced QFI and quantum sensitivity. At higher levels of deformation, the even-cat state outperforms the odd-cat state in terms of QFI. $n_{max} = 30$ was chosen for all computations.} 
    \label{fig:qfivsq}
\end{figure}

\begin{figure*}
  \subfigure[]{
\includegraphics[width=.48\textwidth]{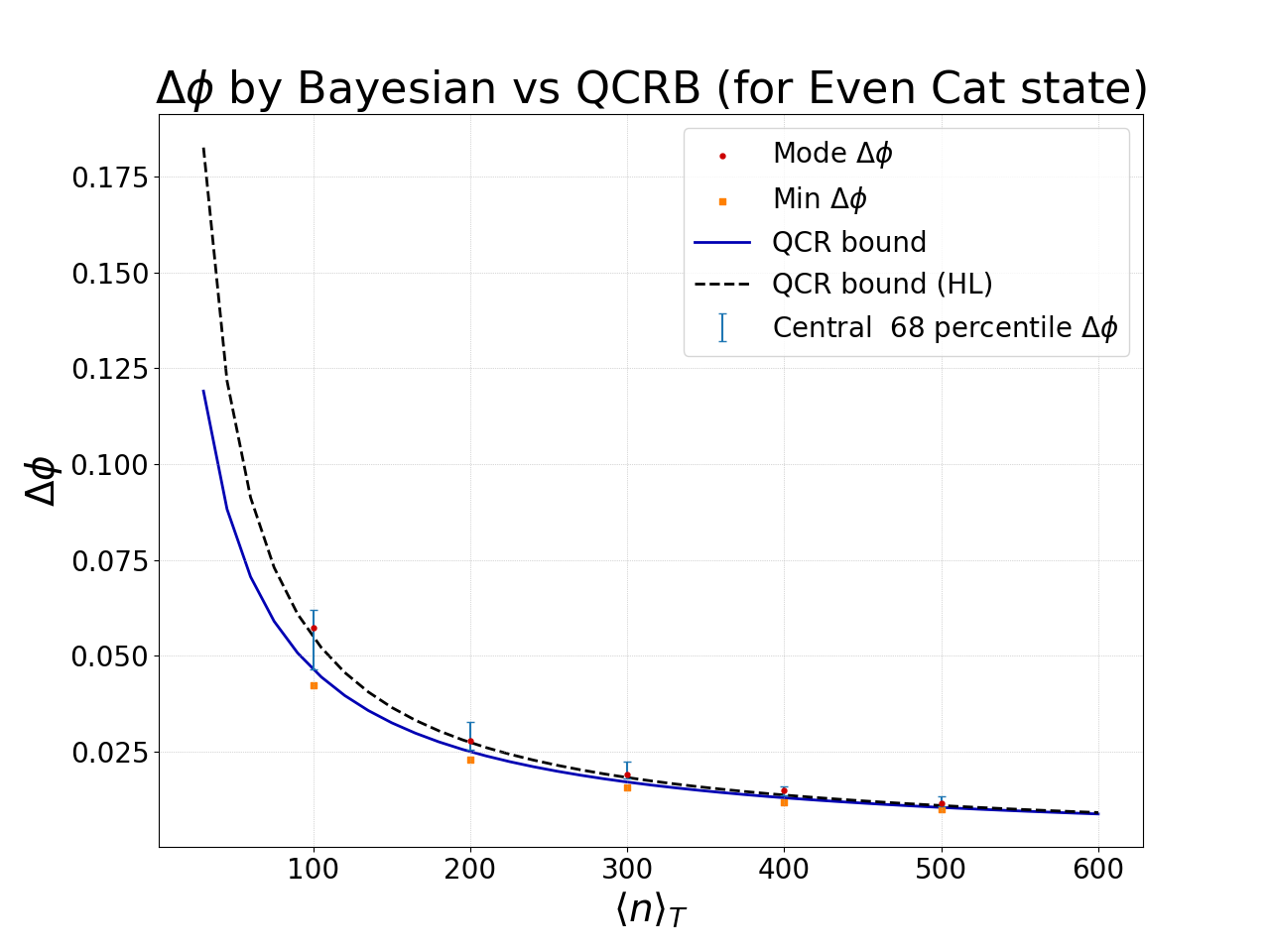}
\label{fig4a}}
\subfigure[]{
   \includegraphics[width=.48\textwidth]{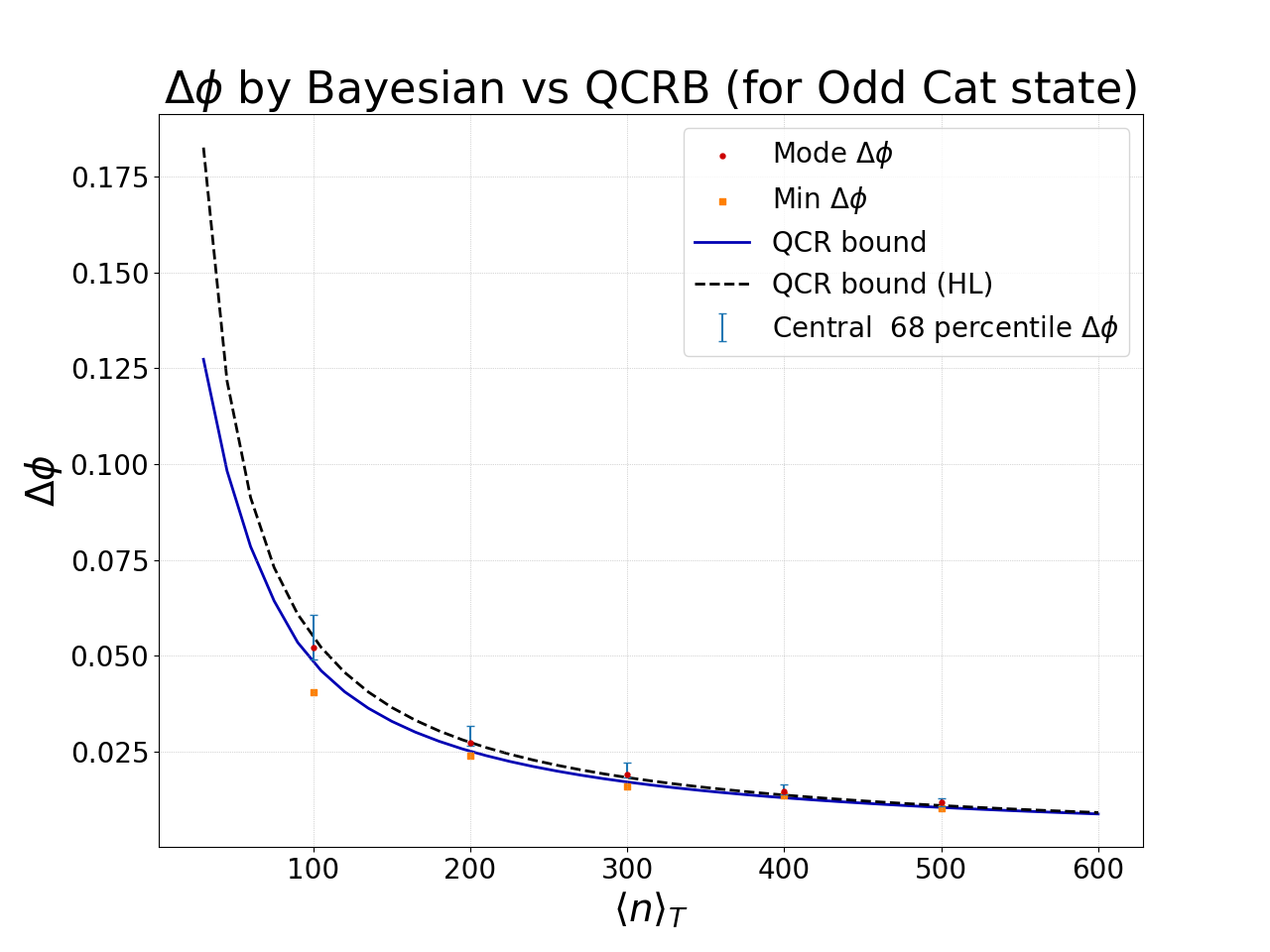}
  \label{fig4b}}
 \caption{(Color online). Blue (solid) line: QCRB on the phase‐uncertainty $\Delta\phi$, computed from the QFI at $q = 0.9$ with a cutoff of $n_{max} = 30$. Black (solid) line: Cram\'er-Rao lower bound given by the HL. Data points: Bayesian estimates of $\Delta \phi$ obtained from simulated detector counts based on $\nu=30$ independent measurements at deformation parameter $q=0.9$ and true phase \(\phi=\pi/2\). For each value of the total mean photon number $\langle n\rangle_T$, 100 simulation runs were performed. Red (circular) markers represent the sample mode of $\Delta\phi$, with error bars indicating the central 68$\%$ credible interval. As $\langle n\rangle_T$ increases, the Bayesian estimates converge toward the Cram\'er-Rao limit, asymptotically saturating the bound.}
\label{fig:Bayesian}
\end{figure*}

Here, we analyze the QFI for two different pairs of $q$-deformed nonlinear input states of the MZI: even cat state $\ket{\text{cat}}_q^+$ with the coherent state $\ket{\alpha}_q$, and odd cat state $\ket{\text{cat}}_q^-$ with the coherent state $\ket{\alpha}_q$. The analytical expressions for the $q$-deformed coherent state $\ket{\alpha}_{q}$ and the cat states $\ket{\text{cat}^{\pm}}_{q}$ are given in Eqs.\,\eqref{qNonlinearCoherent} and \eqref{qCat}, respectively. Since the $q$-deformed ladder operators do not yield closed-form expressions for the Fock-state coefficients or the normalization factors, each state was represented numerically using its series expansion on the photon-number basis.  All series were truncated at a maximum photon number $n_{\max}=30$, beyond which, we verified, that all relevant quantities have sufficiently converged.

In Fig.\,\ref{fig:QFIvsn}, we plot the QFI, given by Eq.\,\eqref{QFIFInal}, as a function of the average input photon number $\langle \hat{n} \rangle$ for both pairs of input states under different values of the deformation parameter $q$. We observe that in both cases, the QFI increases monotonically as the $q$-deformation becomes stronger, i.e., as the parameter $q$ decreases. Specifically, for $q=0.1$ (red dashed line), the QFI is approximately $1.05$ times greater than that of the undeformed case $q=1$ (blue dotted line). This enhancement indicates that the non-Gaussian character introduced by the $q$-deformation leads to a monotonic improvement in metrological performance. All cat-state probes—deformed and undeformed—surpass the HL benchmark (black solid line), and the $q$-deformed cat states yield the highest sensitivity. The QFI of cat probes exceeds the HL benchmark only at low photon numbers; for large $\langle\hat n\rangle$ one finds $F_{Q}(\text{cat}) = \langle\hat n\rangle^{2} + \mathcal{O}\!\bigl(\langle\hat n\rangle\bigr)$~\cite{Munro_Spiller_2011}. So, the scaling of the phase uncertainty ultimately converges with the QCRB imposed by the HL, as can be seen from Fig.\,\ref{fig:Bayesian}. So, we see that for a lower number of measurements or lower expected photon numbers, the cat states perform better than the HL using the same amount of resources. 

In Fig.\,\ref{fig:qfivsq}, we present a direct comparison of the QFI as a function of the deformation parameter $q$ for the even and odd $q$-deformed cat states at a fixed average photon number $\langle n\rangle = 20$. We notice that both states exhibit comparable levels of metrological enhancement values down to approximately $q\approx0.8$. Below this threshold, however, the even cat outperforms its odd counterpart. 

Fig.\,\ref{fig:Bayesian} compares the Bayesian phase uncertainty $\Delta\phi_{B}$ (represented by discrete points) with the QCRB $\Delta\phi=1/\sqrt{\nu F_{C}}$ (shown in solid blue line), for input states comprising a $q$-deformed Schr\"odinger cat state, either even (Fig.\,\ref{fig4a}) or odd (Fig.\,\ref{fig4b}), combined with a $q$-deformed coherent state. The deformation parameter is fixed at $q=0.9$. For each mean photon number $\langle n\rangle$, we performed $\nu = 30$ independent measurements, estimated $\Delta\phi_{B}$ using a Bayesian filter, and repeated the entire procedure 100 times. Red dots indicate the half-sample mode of the resulting 100 values of $\Delta \phi$, while the vertical bars represent the central 68$\%$ credible interval. Orange square markers denote the minimum value of $\Delta \phi$ obtained across the 100 runs. 

The QCRB constrains the uncertainty in the estimated parameter over \textit{all} possible data records, whereas each experimental run yields only a \textit{single} record consisting of $\nu=30$ photon‐count outcomes. As a result, the posterior width~$\Delta\phi_{B}$ fluctuates from run to run and does not necessarily equal $1/\sqrt{\nu F_{C}}$ or, equivalently, $1/\sqrt{\nu F_{Q}}$. As either the number of runs $\nu$ or the total average number $\langle n \rangle_{T}$ increases, these run-to-run fluctuations diminish due to central‐limit scaling. Consequently, the \textit{mode} of the posterior widths over 100 runs approaches the QCRB, even though the upper tail of the distribution remains visible in the 68$\%$ error bars.

The QCRB is strictly asymptotic, emerging from a series expansion that retains only the leading-order term in the inverse number of experimental repetitions, $\nu^{-1}$. Incorporating the higher-order terms in this expansion leads to a corrected lower bound on the mean-square error (MSE)
\begin{align}
 (\Delta \phi)^2 &\ge \frac{1}{\nu} \frac{1}{F_{C}(\phi)} + \frac{1}{\nu^2}\left(\frac{-1}{F_{C}(\phi)} + \gamma_{1} \frac{1}{F^3_{C}(\phi)} + \gamma_{2} \frac{1}{F^4_{C}(\phi)} \right) \notag \\
 &\qquad + O\left(\frac{1}{\nu^3}\right),
\end{align}
where $\nu$ denotes the number of independent measurements, $F_{C}(\phi)$ is the classical Fisher information evaluated at the true parameter value $\phi$, and $\gamma_{1}$ and $\gamma_{2}$ are curvature coefficients determined by higher-order score cumulants~\cite{Hervas_etal_2025}. 

The second-order correction is negative in the first order of $1/{F_{C}(\phi)}$, further lowering the bound on the estimation error. For small values of $F_{C}(\phi)$, the $\nu^{-2}$ corrections can become significant, potentially dominating the behavior and leading to apparent violations of the first-order quantum QCRB. As a result, some of the minimum values of $\Delta \phi$ fall below the QCRB. However, as $F_{C}(\phi)$ increases, specifically, when $F_{C}(\phi) \gg 1$, these higher-order contributions become negligible, and the full bound converges to the well-known asymptotic form, eliminating any such violations. Importantly, even for $q$-deformed states, the Bayesian filter saturates the corrected Cram\'er–Rao bound, or equivalently, the QCRB.
\section{Conclusions and outlook} \label{sec6}
In this work, we have analyzed the phase estimation capabilities of $q$-deformed coherent and Schr\"odinger cat states in an MZI setup. By modeling input states via the $q$-deformation of the bosonic algebra, we captured realistic deviations from ideal photon statistics. Using the Jordan–Schwinger mapping, we derived closed-form photon count distributions and demonstrated that Bayesian inference applied to photon-number-resolving measurements saturates the QCRB across all tested deformations.

Our results reveal two key insights: (i) photon counting remains an optimal measurement strategy for $q$-deformed states, and (ii) the QFI increases with the degree of deformation, indicating that the nonclassical and non-Gaussian features introduced by $q$-deformation can enhance metrological performance. Notably, even modest values of deformation lead to measurable gains in precision, positioning $q$-deformed states as viable and tunable resources for quantum-enhanced sensing. These findings, supported by theoretical proposals \cite{Wang_Goorskey_Xiao_2001, Naderi_Soltanolkotabi_Roknizadeh_2005, Lu_etal_2013, Yan_Zhu_Li_2016} and experimental demonstrations \cite{Gambetta_etal_2006, Solntsev_etal_2014, Latmiral_Mintert_2018, Kruk_etal_2019} of nonlinear nonclassical states, underscore their feasibility and growing importance in advancing quantum sensing and information processing technologies.

These findings open several avenues for future exploration. Building on the existing experimental realization of nonlinear nonclassical states, efforts can now focus on the generation and characterization of $q$-deformed states in controllable platforms, such as engineered media or feedback-stabilized lasers, for deployment in practical quantum metrology platforms. Theoretically, extending the analysis to multi-parameter estimation, adaptive protocols, and other classes of deformed algebras may further elucidate the advantages of algebraic deformation in quantum sensing. Additionally, understanding the interplay between deformation, decoherence, and error correction could inform the design of robust quantum sensors for next-generation precision technologies.\\ \vspace{0.1cm}

\noindent \textbf{\large{Acknowledgments:}} Duttatreya is supported by a UGC-NET PhD Research Scholarship. S.D. thanks S. Omkar for useful discussions and acknowledges the support of research grants DST/FFT/NQM/QSM/2024/3 (by DST-National Quantum Mission, Govt. of India), and NFSG/GOA/2023/G0928 (by BITS-Pilani).



\end{document}